\documentclass[12pt]{article}
\usepackage{latexsym,amssymb,amsmath}
\begin{document}
\baselineskip=18pt

\newcommand{\la}{\langle}
\newcommand{\ra}{\rangle}
\newcommand{\psp}{\vspace{0.4cm}}
\newcommand{\pse}{\vspace{0.2cm}}
\newcommand{\ptl}{\partial}
\newcommand{\dlt}{\delta}
\newcommand{\sgm}{\sigma}
\newcommand{\al}{\alpha}
\newcommand{\be}{\beta}
\newcommand{\G}{\Gamma}
\newcommand{\gm}{\gamma}
\newcommand{\vs}{\varsigma}
\newcommand{\Lmd}{\Lambda}
\newcommand{\lmd}{\lambda}
\newcommand{\td}{\tilde}
\newcommand{\vf}{\varphi}
\newcommand{\yt}{Y^{\nu}}
\newcommand{\wt}{\mbox{wt}\:}
\newcommand{\rd}{\mbox{Res}}
\newcommand{\ad}{\mbox{ad}}
\newcommand{\stl}{\stackrel}
\newcommand{\ol}{\overline}
\newcommand{\ul}{\underline}
\newcommand{\es}{\epsilon}
\newcommand{\dmd}{\diamond}
\newcommand{\clt}{\clubsuit}
\newcommand{\vt}{\vartheta}
\newcommand{\ves}{\varepsilon}
\newcommand{\dg}{\dagger}
\newcommand{\tr}{\mbox{Tr}}
\newcommand{\ga}{{\cal G}({\cal A})}
\newcommand{\hga}{\hat{\cal G}({\cal A})}
\newcommand{\Edo}{\mbox{End}\:}
\newcommand{\for}{\mbox{for}}
\newcommand{\kn}{\mbox{ker}}
\newcommand{\Dlt}{\Delta}
\newcommand{\rad}{\mbox{Rad}}
\newcommand{\rta}{\rightarrow}
\newcommand{\mbb}{\mathbb}
\newcommand{\lra}{\Longrightarrow}

\begin{center}{\Large \bf   Stable-Range Approach to Short  Wave
 }\end{center}
\begin{center}{\Large \bf   and Khokhlov-Zabolotskaya  Equations}
\end{center}
\vspace{0.2cm}

\begin{center}{\large Xiaoping Xu}\end{center}
\begin{center}{Institute of Mathematics, Academy of Mathematics \& System Sciences}\end{center}
\begin{center}{Chinese Academy of Sciences, Beijing 100190, P.R. China}
\footnote{Research supported
 by China NSF 10871193}\end{center}

\vspace{0.6cm}

 \begin{center}{\Large\bf Abstract}\end{center}

\vspace{1cm} {\small Short wave equations were introduced in
connection with the nonlinear reflection of weak shock waves. They
also relate to the modulation of a gas-fluid mixture.
Khokhlov-Zabolotskaya equation are used to describe the propagation
of a diffraction sound beam in a nonlinear medium. We give a new
algebraic method of solving these equations by using certain
finite-dimensional stable range of the nonlinear terms and  obtain
large families of new explicit exact solutions parameterized by
several functions for them. These parameter functions enable one to
find the solutions of some related practical models and boundary
value problems.}

\vspace{0.8cm}

\section{Introduction}

 Khristianovich and Rizhov [8] (1958) discovered the
equations of short waves in connection with the nonlinear reflection
of weak shock waves. The equations are mathematically equivalent to
the following equation of their potential function $u$ for the
velocity vector:
$$2u_{tx}-2(x+u_x)u_{xx}+u_{yy}+2ku_x=0,\eqno(1.1)$$
 where $k$ is a real constant. For convenience, we call the above
 equation ``the short wave equation".  The
symmetry group and conservation laws of (1.1) were first studied by
Kucharczyk [14] (1965) and later by Khamitova [6] (1982).
 Bagdoev and Petrosyan [2] (1985)
showed that the modulation equation of a gas-fluid mixture coincides
in main orders with the corresponding short-wave equation. Roy, Roy
and De [23] (1988) found a loop algebra in the Lie symmetries for
the short-wave equation. Kraenkel, Manna and Merle [13]  (2000)
studied nonlinear short-wave propagation in ferrites and Ermakov [3]
(2006) investigated short-wave interaction in film slicks.

Khokhlov and Zabolotskaya [7] (1969) found the equation
$$2u_{tx}+(uu_x)_x-u_{yy}=0.\eqno(1.2)$$
for quasi-plane waves in nonlinear acoustics of bounded bundles.
More specifically,  the equation describes the propagation of a
diffraction sound beam in a nonlinear medium (cf. [4], [20]).
Kupershmidt [15] (1994) constructed a geometric Hamiltonian form for
the Khokhlov-Zabolotskaya equation (1.2). Certain group-invariant
solutions of (1.2) were found by Korsunskii [12] (1991), and by Lin
and Zhang [16] (1995).
 The three-dimensional generalization
$$2u_{tx}+(uu_x)_x-u_{yy}-u_{zz}=0\eqno(1.3)$$
and  its symmetries were studied by  Krasil'shchik, Lychagin and
Vinogradov [17] (1986) and by Schwarz [25] (1987). Martinez-Moras
and Ramos [18] (1993) showed that the higher dimensional classical
W-algebras are the Poisson structures associated with a higher
dimensional version of the Khokhlov-Zabolotskaya hierarchy.
 Kacdryavtsev and Sapozknikov [9] (1998) found the
symmetries for a generalized Khokhlov-Zabolotskaya equation. Sanchez
[24] (2007) studied long waves in ferromagnetic media via
Khokhlov-Zabolotskaya equation.
 Morozov [19]
(2008) derived two non-equivalent coverings for the modified
Khokhlov-Zabolotskaya equation from Maurer-Cartan forms of its
symmetry pseudo-group. Rozanova [21, 22] (2007, 2008) studied
closely related Khokhlov- Zabolotskaya-Kuzentsov equation from
analytic point of view. Kostin and Panasenko [11] (2008)
investigated nonlinear acoustics in heterogeneous media via
Khokhlov- Zabolotskaya-Kuzentsov-type equation. All the above
equations are similar nonlinear algebraic partial differential
equations.

Observe that the nonlinear terms in the above equations keep some
finite-dimensional polynomial space in $x$ stable. In this paper, we
present a new algebraic method of solving these equations by using
this stability. We obtain a family of solutions of the equation
(1.1) with $k=1/2,2$, which blow up on a moving line $y=f(t)$. They
may reflect partial phenomena of gust. Moreover, we obtain another
family of smooth solutions parameterized by six smooth functions of
$t$ for any $k$. Similar results for the equation (1.2) are also
given. Furthermore, we find a family of solutions of the equation
(1.3) blowing up on a rotating and translating plane
$\cos\al(t)\:y+\sin\al(t)\:z=f(t)$, which may reflect partial
phenomena of sound shock, and a family of solutions
 parameterized by time-dependent harmonic functions
in $y$ and $z$, whose special cases are smooth solutions. Since our
solutions contain parameter functions, they can be used to solve
certain related practical models and  boundary-value problems for
these equations.

 On the list of the  Lie point symmetries of the equation
(1.1) in the works of Kucharczyk [14] and of Khamitova [6] (e.g. cf.
Page 301 in [5]), the most sophisticated ones are those with respect
to the following vector fields:
$$X_1=-\al'y\ptl_x+\al\ptl_y+\left[xy({\al'}'+\al')-\frac{y^3}{3}({{\it
\al'}'}'+(k+1){\al'}'+k\al')\right]\ptl_u,\eqno(1.4)$$
$$X_2=\be\ptl_x+[y^2({\be'}'+(k+1)\be'+k\be)-x(\be'+\be)]\ptl_u,\eqno(1.5)$$ where $\al$
 and
$\be$ are arbitrary functions of $t$. Among the known Lie point
symmetries of the Khokhlov-Zabolotskaya equation (1.2) in the works
of Vinogradov and Vorob'ev [26], and of Schwarz [25]
 (e.g. cf. Page 299 in [5]), the most interesting ones are those with respect
to the following vector fields:
$$X_3=\frac{1}{2}\al'y\ptl_x+\al\ptl_y-\frac{1}{2}{\al'}'y\ptl_u,\eqno(1.6)$$
$$X_4=\be\ptl_t+\frac{2\be'x+{\be'}'y^2}{6}\ptl_x+\frac{2}{3}\be'
y\ptl_y-\frac{4\be'u+2{\be'}'x+{{\be'}'}'y^2}{6}\ptl_u. \eqno(1.7)$$
The symmetries of the three-dimensional Khokhlov-Zabolotskaya
equation (1.3) causing our attention are those with respect to the
vector fields (e.g. cf. Page 301 in [5]):
$$X_5=10t^2\ptl_t+(4tx+3y^2+3z^2)\ptl_x+12ty\ptl_y+12tz\ptl_z-(4x+16tu)\ptl_u,\eqno(1.7)$$
$$X_6=\frac{1}{2}\al'y\ptl_x+\al\ptl_z-\frac{1}{2}{\al'}'y\ptl_u,\eqno(1.8)$$
$$X_7=\frac{1}{2}\be'z\ptl_x+\be\ptl_y-\frac{1}{2}{\be'}'z\ptl_u.\eqno(1.9)$$

We find that the group-invariant solutions with respect to the
above vector fields $X_1$-$X_7$ are polynomial in $x$. This
motivates us to find more exact solutions of the equations
(1.1)-(1.3) polynomial in $x$.

In Section 2, we solve the short-wave equation (1.1). Although the
equation (1.2) can be viewed as a special case of the equation
(1.3), we first solve (1.2) in Section 3 for simplicity because our
approach to (1.3) involves time-dependent harmonic functions and
sophisticated integrals. The exact solutions of the equation (1.3)
will be given Section 4.

\section{Short Wave Equation}

In this section, we study solutions polynomial in $x$ for the
short wave equation (1.1).
 By comparing the terms of highest degree in $x$, we find that
 such a solution must be of the form:
 $$u=f(t,y)+g(t,y)x+h(t,y)x^2+\xi(t,y)x^3,\eqno(2.1)$$
where $f(t,y),\;g(t,y),\;h(t,y)$ and $\xi(t,y)$ are
suitably-differentiable functions to be determined. Note
$$u_x=g+2hx+3\xi x^2,\qquad u_{xx}=2h+6\xi x,\eqno(2.2)$$
$$u_{tx}=g_t+2h_tx+3\xi_t x^2,\qquad
u_{yy}=f_{yy}+g_{yy}x+h_{yy}x^2+\xi_{yy}x^3,\eqno(2.3)$$ Now (1.1)
becomes \begin{eqnarray*}\hspace{1cm}& & 2(g_t+2h_tx+3\xi_t
x^2)-2(g+(2h+1)x+3\xi x^2)(2h+6\xi x)\\ & &
+f_{yy}+g_{yy}x+h_{yy}x^2+\xi_{yy}x^3+2k(g+2hx+3\xi
x^2)=0,\hspace{3.5cm}(2.4)\end{eqnarray*} which is equivalent to the
following systems of partial differential equations:
$$\xi_{yy}=36\xi^2,\eqno(2.5)$$
$$h_{yy}=6\xi(6h+2-k)-6\xi_t,\eqno(2.6)$$
$$g_{yy}=8h^2+4(1-k)h+12\xi g-4h_t,\eqno(2.7)$$
$$f_{yy}=4gh-2g_t-2kg.\eqno(2.8)$$

First we observe that
$$\xi=\frac{1}{(\sqrt{6}y+\be(t))^2}\eqno(2.9)$$ is a solution of
the equation (2.5) for any differentiable function $\be$ of $t$.
Substituting (2.9) into (2.6), we get
$$h_{yy}=\frac{12\be'(t)}{(\sqrt{6}y+\be(t))^3}+\frac{6(6h+2-k)}{(\sqrt{6}y+\be(t))^2}.
\eqno(2.10)$$ Denote by $\mbb{Z}$ the ring of integers. Write
$$h(t,y)=\sum_{i\in\mbb{Z}}a_i(t)(\sqrt{6}y+\be(t))^i.\eqno(2.11)$$
Then
$$h_{yy}=\sum_{i\in\mbb{Z}}6(i+2)(i+1)a_{i+2}(t)(\sqrt{6}y+\be(t))^i.\eqno(2.12)$$
Substituting (2.11) and (2.12) into (2.10), we obtain
$$\sum_{i\in\mbb{Z}}6[(i+2)(i+1)-6]a_{i+2}(t)(\sqrt{6}y+\be(t))^i=
\frac{12\be'(t)}{(\sqrt{6}y+\be(t))^3}+\frac{6(2-k)}{(\sqrt{6}y+\be(t))^2}.
\eqno(2.13)$$ So
$$-24a_{-1}(t)=12\be'(t),\qquad -36a_0(t)=6(2-k)\eqno(2.14)$$
and
$$6(i+4)(i-1)a_{i+2}(t)=0,\qquad i\neq -2,-3.\eqno(2.15)$$ Thus
$$h=\frac{\al}{(\sqrt{6}y+\be)^2}-\frac{\be'}{2(\sqrt{6}y+\be)}+\frac{k-2}{6}
+\gm(\sqrt{6}y+\be)^3,\eqno(2.16)$$ where $\al$ and $\gm$ are
arbitrary  differentiable functions of $t$.

Note
$$h_t=\frac{-2\al\be'}{(\sqrt{6}y+\be)^3}+
\frac{2\al'+(\be')^2}{2(\sqrt{6}y+\be)^2}-\frac{{\be'}'}{2(\sqrt{6}y+\be)}
+3\gm\be' (\sqrt{6}y+\be)^2+\gm' (\sqrt{6}y+\be)^3\eqno(2.17)$$
and
\begin{eqnarray*}& &h^2=\frac{\al^2}{(\sqrt{6}y+\be)^4}-
\frac{\al\be'}{(\sqrt{6}y+\be)^3}+\frac{3(\be')^2+4(k-2)\al}{12(\sqrt{6}y+\be)^2}
+\frac{(2-k)\be'}{6(\sqrt{6}y+\be)}+\frac{(k-2)^2}{36}
\\ &
&+2\al\gm(\sqrt{6}y+\be)-\be'\gm(\sqrt{6}y+\be)^2+\frac{(k-2)\gm}{3}(\sqrt{6}y+\be)^3+\gm^2(\sqrt{6}y+\be)^6.
\hspace{1.1cm}(2.18)\end{eqnarray*} Substituting the above two
equations into (2.7), we have:
\begin{eqnarray*}&&
g_{yy}-\frac{12g}{(\sqrt{6}y+\be)^2}
 =\frac{8\al^2}{(\sqrt{6}y+\be)^4}
-\frac{4[(k+1)\al+3\al']}{3(\sqrt{6}y+\be)^2}+\frac{2((k+1)\be'+3{\be'}')}{3(\sqrt{6}y+\be)}
\hspace{4cm}\end{eqnarray*}
\begin{eqnarray*}
&&+\frac{2(k-2)(1-2k)}{9}
+16\al\gm(\sqrt{6}y+\be) -20\be'\gm(\sqrt{6}y+\be)^2\\
\\ &&-\frac{4[(k+1)\gm+3\gm']}{3}(\sqrt{6}y+\be)^3+8\gm^2(\sqrt{6}y+\be)^6.
\hspace{5.5cm}(2.19)\end{eqnarray*} Write
$$g(t,y)=\sum_{i\in\mbb{Z}}b_i(t)(\sqrt{6}y+\be)^i.\eqno(2.20)$$
Then
\begin{eqnarray*}\hspace{2cm}&&\sum_{i\in\mbb{Z}}6[(i+2)(i+1)-2]b_{i+2}(t)(\sqrt{6}y+\be)^i
\\&=&\frac{8\al^2}{(\sqrt{6}y+\be)^4}
-\frac{4[(k+1)\al+3\al']}{3(\sqrt{6}y+\be)^2}
+\frac{2[(k+1)\be'+3{\be'}']}{3(\sqrt{6}y+\be)}\\
&&+\frac{2(k-2)(1-2k)}{3} +16\al\gm(\sqrt{6}y+\be)
-20\be'\gm(\sqrt{6}y+\be)^2\\ & &
-\frac{4[(k+1)\gm+3\gm']}{3}(\sqrt{6}y+\be)^3+8\gm^2(\sqrt{6}y+\be)^6.
\hspace{3.2cm}(2.21)\end{eqnarray*} Comparing the constant terms,
we get $k=1/2,\;2$. Moreover, the coefficients of the other terms
give
$$b_{-2}=\frac{\al^2}{3},\;\;b_0=\frac{(k+1)\al+3\al'}{9},\;\;
b_1=-\frac{(k+1)\be'+3{\be'}'}{18},\eqno(2.22)$$
$$b_3=\frac{2\al\gm}{3},\;\;b_4=-\frac{\be'\gm}{3},\;\;b_5=-\frac{(k+1)\gm+3\gm'}
{81},\;\; b_8=\frac{2\gm^2}{81}\eqno(2.23)$$ and
$$(i+3)ib_{i+2}=0\qquad \for\;\;i\neq -4,-2,-1,1,2,3,6.\eqno(2.24)$$
Therefore
\begin{eqnarray*}g&=&\frac{\al^2}{3(\sqrt{6}y+\be)^2}+\frac{\sgm}{\sqrt{6}y+\be}
+\frac{(k+1)\al+3\al'}{9}-\frac{(k+1)\be'+3{\be'}'}{18}
(\sqrt{6}y+\be)\\
& &+\rho(\sqrt{6}y+\be)^2 +\frac{2\al\gm}{3}(\sqrt{6}y+\be)^3
-\frac{\be'\gm}{3}(\sqrt{6}y+\be)^4\\ & & -\frac{(k+1)\gm+3\gm'}
{81}(\sqrt{6}y+\be)^5+\frac{2\gm^2}{81}
(\sqrt{6}y+\be)^8,\hspace{5.3cm}(2.25)\end{eqnarray*} where $\sgm$
and $\rho$ are arbitrary differentiable functions of $t$.

Observe that
\begin{eqnarray*}& &g_t=-\frac{2\al^2\be'}{3(\sqrt{6}y+\be)^3}+
\frac{(2\al\al'-3\sgm\be')}{3(\sqrt{6}y+\be)^2}+\frac{\sgm'}{\sqrt{6}y+\be}
+\frac{(k+1)(2\al'-(\be')^2)}{18}\\
&
&+\frac{2{\al'}'-\be'{\be'}'}{6}+\frac{36\be'\rho-(k+1){\be'}'-3{{\be'}'}'}{18}
(\sqrt{6}y+\be)+(\rho'+2\al\be'\gm)(\sqrt{6}y+\be)^2
\\
& & +\frac{2\al\gm'+2\al'\gm-4(\be')^2\gm}{3}(\sqrt{6}y+\be)^3
-\frac{42\be'\gm'+5(k+1)\be'\gm+27{\be'}'\gm}{81}(\sqrt{6}y+\be)^4 \\
&&-\frac{(k+1)\gm'+3{\gm'}'} {81}
(\sqrt{6}y+\be)^5+\frac{16\be'\gm^2}{81} (\sqrt{6}y+\be)^7
+\frac{4\gm\gm'}{81}
(\sqrt{6}y+\be)^8,\hspace{1.7cm}(2.26)\end{eqnarray*}
\begin{eqnarray*}&
&gh=\frac{\al^3}{3(\sqrt{6}y+\be)^4}+\frac{6\al\sgm-\al^2\be'}{6(\sqrt{6}y+\be)^3}
+\frac{k\al^2+2\al\al'-3\be'\sgm}{6(\sqrt{6}y+\be)^2}
\\ & &+\frac{3(k-2)\sgm-3\al{\be'}'-2(k+1)\al\be'-3\al'\be'}{18(\sqrt{6}y+\be)}
+\al\rho+\frac{(k+1)(\be')^2+3\be'{\be'}'}{36}\\ &
&+\frac{(k-2)((k+1)\al+3\al')}{54} +[\al^2\gm-\frac{\be'\rho}{2}-
\frac{(k-2)((k+1)\be'+3{\be'}')}{108}](\sqrt{6}y+\be)
\\ & &+\frac{(k-2)\rho+6\gm\sgm-4\al\be'\gm}{6}(\sqrt{6}y+\be)^2
+[\frac{(\be')^2\gm}{6}+\frac{(17k-10)\al\gm-3\al\gm'}{81}\\&&+\frac{\al'\gm}{3}](\sqrt{6}y+\be)^3+\frac{(10-17k)\be'\gm+3\be'\gm'-27{\be'}'\gm}{162}
(\sqrt{6}y+\be)^4\\ & &+ [\gm\rho-\frac{(k-2)((k+1)\gm+3\gm')}
{486}](\sqrt{6}y+\be)^5 +\frac{56\al\gm^2}{81} (\sqrt{6}y+\be)^6
\\ & &-\frac{28\be'\gm^2}{81}
 (\sqrt{6}y+\be)^7+\frac{(k-2)\gm^2}{243}
(\sqrt{6}y+\be)^8 +\frac{2\gm^3}{81}
(\sqrt{6}y+\be)^{11}.\hspace{2.7cm}(2.27)\end{eqnarray*}
Substituting (2.25), (2.26) and (2.27) into (2.8), we obtain
\begin{eqnarray*}&
&f_{yy}=\frac{4\al^3}{3(\sqrt{6}y+\be)^4}
-\frac{2[6(k+1)\sgm+3\al{\be'}'+2(k+1)\al\be'+3\al'\be'+9\sgm']}{9(\sqrt{6}y+\be)}
\hspace{7cm}\end{eqnarray*}
\begin{eqnarray*}  &
&+\frac{12\al\sgm+2\al^2\be'}{3(\sqrt{6}y+\be)^3}+4\al\rho+
\frac{2[\be'{\be'}'-(k+1)\al'-{\al'}']}{3}+\frac{2(k+1)(\be')^2}{9}\\
& &-\frac{4(k+1)^2\al}{27}
+[4\al^2\gm-6\be'\rho+\frac{{{\be'}'}'+(k+1){\be'}'}{3}+\frac{2(k+1)^2\be'}{27}]
(\sqrt{6}y+\be)\\
&&+[4\gm\sgm-2\rho'-\frac{4(k+1)\rho+20\al\be'\gm}{3}](\sqrt{6}y+\be)^2
+[-\frac{40((k+1)\al\gm+3\al\gm')}{81}\\ &
&+\frac{10(\be')^2\gm}{3}](\sqrt{6}y+\be)^3
+\frac{10(k+1)\be'\gm+30\be'\gm'}{27}(\sqrt{6}y+\be)^4+[4\gm\rho\\
& &+\frac{4(k+1)^2\gm}{243}
+\frac{2(k+1)\gm'+2{\gm'}'}{27}](\sqrt{6}y+\be)^5
+\frac{224\al\gm^2}{81} (\sqrt{6}y+\be)^6
\\&&-\frac{16\be'\gm^2}{9}
(\sqrt{6}y+\be)^7-\frac{8\gm((k+1)\gm+3\gm')}{243}
(\sqrt{6}y+\be)^8 +\frac{8\gm^3}{81}
(\sqrt{6}y+\be)^{11}.\hspace{1.1cm}(2.28)\end{eqnarray*} Thus
\begin{eqnarray*}&
&f=\frac{\al^3}{27(\sqrt{6}y+\be)^2}+\frac{6\al\sgm+\al^2\be'}{18(\sqrt{6}y+\be)}+\theta+\vt
y+2\al\rho y^2+\frac{(k+1)(\be')^2}{9}y^2\\
&&-\frac{6(k+1)\sgm+3\al{\be'}'+2(k+1)\al\be'+3\al'\be'+9\sgm'}{27}(\sqrt{6}y+\be)[\ln
(\sqrt{6}y+\be)-1]\\ & &+
\frac{\be'{\be'}'-(k+1)\al'-{\al'}'}{3}y^2-\frac{2(k+1)^2\al}{27}y^2
+[\frac{\al^2\gm}{9}-\frac{\be'\rho}{6}
+\frac{{{\be'}'}'+(k+1){\be'}'}{108}\\&& +\frac{(k+1)^2\be'}{486}]
(\sqrt{6}y+\be)^3+[\frac{2\gm\sgm-\rho'}{36}-\frac{(k+1)\rho+5\al\be'\gm}{54}](\sqrt{6}y+\be)^4
\\
&&+[\frac{(\be')^2\gm}{36}-\frac{(k+1)\al\gm+3\al\gm'}{243}](\sqrt{6}y+\be)^5
+\frac{(k+1)\be'\gm+3\be'\gm'}{486}(\sqrt{6}y+\be)^6\\ &
&+[\frac{\gm\rho}{63}+\frac{(k+1)^2\gm}{15309}
+\frac{(k+1)\gm'+{\gm'}'}{3402}](\sqrt{6}y+\be)^7
+\frac{2\al\gm^2}{243} (\sqrt{6}y+\be)^8
\\&&-\frac{\be'\gm^2}{243}
(\sqrt{6}y+\be)^9-\frac{2\gm((k+1)\gm+3\gm')}{32805}
(\sqrt{6}y+\be)^{10} +\frac{\gm^3}{9477}
(\sqrt{6}y+\be)^{13},\hspace{1.1cm}(2.29)\end{eqnarray*}
 where
$\theta$ and $\vartheta$ are arbitrary functions of $t$. \psp

{\bf Theorem 2.1}. {\it When $k=1/2,\;2$, we have the following
solution of the equation (1.1) blowing up on the surface
$\sqrt{6}y+\be(t)=0$:
\begin{eqnarray*}&
&u=\frac{x^3}{(\sqrt{6}y+\be)^2}+[\frac{\al}{(\sqrt{6}y+\be)^2}-\frac{\be'}{2(\sqrt{6}y+\be)}
+\frac{k-2}{6} +\gm(\sqrt{6}y+\be)^3]x^2\\ & &
+[\frac{\al^2}{3(\sqrt{6}y+\be)^2}+\frac{\sgm}{\sqrt{6}y+\be}
+\frac{(k+1)\al+3\al'}{9}-\frac{(k+1)\be'+3{\be'}'}{18}
(\sqrt{6}y+\be)\\
& &+\rho(\sqrt{6}y+\be)^2 +\frac{2\al\gm}{3}(\sqrt{6}y+\be)^3
-\frac{\be'\gm}{3}(\sqrt{6}y+\be)^4 -\frac{(k+1)\gm+3\gm'}
{81}(\sqrt{6}y+\be)^5\\& &+\frac{2\gm^2}{81} (\sqrt{6}y+\be)^8]x+
\frac{\al^3}{27(\sqrt{6}y+\be)^2}+\frac{6\al\sgm+\al^2\be'}{18(\sqrt{6}y+\be)}+2\al\rho y^2+\frac{(k+1)(\be')^2}{9}y^2\\
&&-\frac{6(k+1)\sgm+3\al{\be'}'+2(k+1)\al\be'+3\al'\be'+9\sgm'}{27}(\sqrt{6}y+\be)
[\ln (\sqrt{6}y+\be)-1]\\& &+
\frac{\be'{\be'}'-(k+1)\al'-{\al'}'}{3}y^2-\frac{2(k+1)^2\al}{27}y^2
+[\frac{\al^2\gm}{9}-\frac{\be'\rho}{6}
+\frac{{{\be'}'}'+(k+1){\be'}'}{108}\\ & &
+\frac{(k+1)^2\be'}{486}]
(\sqrt{6}y+\be)^3+[\frac{2\gm\sgm-\rho'}{36}-\frac{(k+1)\rho+5\al\be'\gm}{54}](\sqrt{6}y+\be)^4
+\theta+\vt y\hspace{7cm}\end{eqnarray*}
\begin{eqnarray*}
&&+[\frac{(\be')^2\gm}{36}-\frac{(k+1)\al\gm+3\al\gm'}{243}](\sqrt{6}y+\be)^5
+\frac{(k+1)\be'\gm+3\be'\gm'}{486}(\sqrt{6}y+\be)^6\\
& &+[\frac{\gm\rho}{63}+\frac{(k+1)^2\gm}{15309}
+\frac{(k+1)\gm'+{\gm'}'}{3402}](\sqrt{6}y+\be)^7
+\frac{2\al\gm^2}{243} (\sqrt{6}y+\be)^8
\\&&-\frac{\be'\gm^2}{243}
(\sqrt{6}y+\be)^9-\frac{2\gm((k+1)\gm+3\gm')}{32805}
(\sqrt{6}y+\be)^{10} +\frac{\gm^3}{9477}
(\sqrt{6}y+\be)^{13},\hspace{1.1cm}(2.30)\end{eqnarray*} where
$\al,\be,\gm,\sgm,\rho,\theta$ and $\vartheta$ are arbitrary
functions of $t$, whose derivatives appeared in the above exist in
a certain open set of $\mbb{R}$.}\psp

When $\al=\gm=\sgm=\rho=\theta=\vartheta=0$, the above solution
becomes
\begin{eqnarray*}&
&u=\frac{x^3}{(\sqrt{6}y+\be)^2}+[\frac{k-2}{6}-\frac{\be'}{2(\sqrt{6}y+\be)}]x^2
-\frac{(k+1)\be'+3{\be'}'}{18} (\sqrt{6}y+\be)x
\\ &&+\frac{(k+1)(\be')^2}{9}y^2+ \frac{\be'{\be'}'}{3}y^2
+[\frac{{{\be'}'}'+(k+1){\be'}'}{108}+\frac{(k+1)^2\be'}{486}]
(\sqrt{6}y+\be)^3.\hspace{1.4cm}(2.31)\end{eqnarray*}

Take the trivial solution $\xi=0$ of (2.5), which is the only
solution polynomial in $y$. Then (2.6) and (2.7) become
$$h_{yy}=0,\qquad g_{yy}=8h^2+4(1-k)h-4h_t,\eqno(2.32)$$
Thus
$$h=\al(t)+\be(t)y.\eqno(2.33)$$
Hence
$$g_{yy}=4(2\al^2+(1-k)\al-\al')+4(4\al\be+(1-k)\be-\be')y+8\be^2y^2.\eqno(2.34)$$ So
$$g=\gm+\sgm y+2(2\al^2+(1-k)\al-\al')y^2+\frac{2}{3}(4\al\be+(1-k)\be-\be')y^3+\frac{2}{3}
\be^2y^4,\eqno(2.35)$$ where $\gm$ and $\sgm$ are arbitrary
functions of $t$. Now (2.8) yields
\begin{eqnarray*} f_{yy}&=&
4(\al+\be y)[\gm+\sgm
y+2(2\al^2+(1-k)\al-\al')y^2+\frac{2}{3}(4\al\be+(1-k)\be-\be')y^3\\
& & +\frac{2}{3} \be^2y^4]-2[\gm'+\sgm'
y+2(4\al\al'+(1-k)\al'-{\al'}')y^2+\frac{2}{3}(4\al'\be+4\al\be'\\
& & +(1-k)\be'-{\be'}')y^3 +\frac{4}{3} \be\be'y^4]-2k[\gm+\sgm
y+2(2\al^2+(1-k)\al-\al')y^2\\ &&
+\frac{2}{3}(4\al\be+(1-k)\be-\be')y^3+\frac{2}{3} \be^2y^4]\\
&=&4\al\gm-2\gm'-2k\gm+2(2\al\sgm+2\be\gm-\sgm'-k\gm)y+8((1-k)\al-\al')\be
y^3\\&&
+4(4\al^3+2(1-2k)\al^2-6\al\al'+k(k-1)\al+(2k-1)\al'+{\al'}'+\be\sgm)y^2
\\ & &+\frac{4}{3}(20\al^2\be+2(1-3k)\al\be-6\al\be'-4\al'\be+(2k-1)\be'+{\be'}'-k(1-k)\be)y^3
\\
&&+\frac{4}{3}(10\al\be^2+(2-3k)\be^2-4\be\be')y^4+\frac{8}{3}
\be^3y^5.\hspace{5.4cm}(2.36)\end{eqnarray*} Therefore,
\begin{eqnarray*} f&=&
(2\al\gm-\gm'-k\gm)y^2+\frac{2\al\sgm+2\be\gm-\sgm'-k\gm}{3}y^3+\frac{2((1-k)\al-\al')}
{5}y^5+\tau+\rho y\hspace{7cm}\end{eqnarray*}
\begin{eqnarray*} & &
+\frac{1}{3}(4\al^3+2(1-2k)\al^2-6\al\al'+k(k-1)\al+(2k-1)\al'+{\al'}'+\be\sgm)y^4
\\ &&+\frac{1}{15}(20\al^2\be+2(1-3k)\al\be-6\al\be'-4\al'\be+(2k-1)\be'+{\be'}'-k(1-k)\be)y^5
\\&&+\frac{2}{45}(10\al\be^2+(2-3k)\be^2-4\be\be'
)y^6+\frac{4\be^3}{63}y^7.\hspace{5.6cm}(2.37)\end{eqnarray*}
 \pse

{\bf Theorem 2.2}. {\it The following is a solution of the
equation (1.1):
\begin{eqnarray*} & &u=(\al+\be y)x^2+[\gm+\sgm y+2(2\al^2+(1-k)\al-\al')y^2
+\frac{2}{3}(4\al\be+(1-k)\be-\be')y^3\\ & &+\frac{2}{3}
\be^2y^4]x+(2\al\gm-\gm'-k\gm)y^2+\frac{2\al\sgm+2\be\gm-\sgm'-k\gm}{3}y^3+\frac{2((1-k)\al-\al')}
{5}y^5\\& &+\tau+\rho y
+\frac{1}{3}(4\al^3+2(1-2k)\al^2-6\al\al'+k(k-1)\al+(2k-1)\al'+{\al'}'+\be\sgm)y^4
\\ &&+\frac{1}{15}(20\al^2\be+2(1-3k)\al\be-6\al\be'-4\al'\be+(2k-1)\be'+{\be'}'-k(1-k)\be)y^5
\\&&+\frac{2}{45}(10\al\be^2+(2-3k)\be^2-4\be\be'
)y^6+\frac{4\be^3}{63}y^7,\hspace{6cm}(2.38)\end{eqnarray*}
 where
$\al,\be,\gm,\sgm,\rho$ and $\tau$ are arbitrary functions of $t$,
whose derivatives appeared in the above exist in a certain open
set of $\mbb{R}$. Moreover, any solution polynomial in $x$ and $y$
of (1.1) must be of the above form. The above solution is smooth
(analytic) if all $\al,\be,\gm,\sgm,\rho$ and $\tau$ are smooth
(analytic) functions of $t$. }\psp

{\bf Remark 2.3}. In addition to the nonzero solution (2.9) of the
equation (2.5), the other nonzero solutions  are of the form
$$\xi=\wp_\iota(\sqrt{6}y+\be(t)),\eqno(2.39)$$
where $\wp_\iota(w)$ is the Weierstrass's elliptic function such
that
$$\wp'_\iota(w)^2=4(\wp_\iota(w)^3-\iota),\eqno(2.40)$$
and $\iota$ is a nonzero constant and $\be$ is any function of
$t$. When $\be$ is not a constant, the solutions of (2.6)-(2.8)
are extremely complicated. If $\be$ is constant, we can take
$\be=0$ by adjusting $\iota$. Any solution of (2.6)-(2.8) with
$h\neq 0$ is also very complicated. Thus the only simple solution
of the equation (1.1) in this case is
$$u=\wp_\iota(\sqrt{6}y)\:x^3.\eqno(2.41)$$

\section{2-D Khokhlov-Zabolotskaya Equation}

The solution of the equation (1.2) polynomial in $x$ must be of
the form
$$u=f(t,y)+g(t,y)x+\xi(t,y)x^2.\eqno(3.1)$$
Then
$$u_x=g+2\xi x,\qquad u_{tx}=g_t+2\xi_t x,\qquad
u_{yy}=f_{yy}+g_{yy}x+\xi_{yy}x^2,\eqno(3.2)$$
$$(uu_x)_x=\ptl_x(fg+(g^2+2f\xi)x+3g\xi x^2+2\xi^2 x^3)=g^2+2f\xi+6g\xi x+6\xi^2
x^2.\eqno(3.3)$$ Substituting them into (1.2), we get
$$2(g_t+2\xi_t x)+g^2+2f\xi+6g\xi
x+6\xi^2-f_{yy}-g_{yy}x-\xi_{yy}x^2=0,\eqno(3.4)$$ equivalently,
$$\xi_{yy}=6\xi^2,\eqno(3.5)$$
$$g_{yy}-6g\xi=4\xi_t,\eqno(3.6)$$
$$f_{yy}-2f\xi=2g_t+g^2.\eqno(3.7)$$

First we observe that
$$\xi=\frac{1}{(y+\be(t))^2}\eqno(3.8)$$ is a solution of
the equation (3.5) for any differentiable function $\be$ of $t$.
Substituting (3.8) into (3.6), we obtain
$$g_{yy}-\frac{6g}{y+\be(t))^2}=-\frac{8\be'(t)}{(y+\be(t))^3}.\eqno(3.9)$$
Write
$$g(t,y)=\sum_{i\in\mbb{Z}}a_i(t)(y+\be(t))^i.\eqno(3.10)$$
Then (3.9) becomes
$$\sum_{i\in\mbb{Z}}[(i+2)(i+1)-6]a_{i+2}(t)(y+\be(t))^i=-\frac{8\be'(t)}{(y+\be(t))^3}.
\eqno(3.11)$$ Thus
$$a_{-1}=2\be',\qquad(i+4)(i-1)a_{i+2}=0\qquad\for\;\;i\neq
-3.\eqno(3.12)$$ Hence
$$g=\frac{\al(t)}{(y+\be(t))^2}+\frac{2\be'(t)}{y+\be(t)}+\gm(t)(y+\be(y))^3,\eqno(3.13)$$
 where $\al$ and $\gm$ are arbitrary  differentiable
functions of $t$.

Note $$g_t=-\frac{2\al\be'}{(y+\be)^3}+
\frac{\al'-2(\be')^2}{(y+\be)^2} +\frac{2{\be'}'}{y+\be}+3\gm\be'
(y+\be)^2+\gm'(\sqrt{3}y+\be)^3\eqno(3.14)$$ and
$$g^2=\frac{\al^2}{(y+\be)^4}+
\frac{4\al\be'}{(y+\be)^3}+\frac{4(\be')^2}{(y+\be)^2}+2\al\gm(y+\be)+4\gm\be'(y+\be)^2+
\gm^2(y+\be)^6.\eqno(3.15)$$ Substituting the above two equations
into (3.7), we have:
\begin{eqnarray*}\hspace{1cm}f_{yy}-\frac{2f}{(y+\be)^2}
 &=&\frac{\al^2}{(y+\be)^4}+
\frac{2\al'}{(y+\be)^2}+\frac{4{\be'}'}{y+\be} +2\al\gm(y+\be)\\ &
&+10\gm\be'(y+\be)^2 +2\gm'
(y+\be)^3+\gm^2(y+\be)^6.\hspace{2.5cm}(3.16)\end{eqnarray*} Write
$$f(t,y)=\sum_{i\in\mbb{Z}}b_i(t)(y+\be)^i.\eqno(3.17)$$
Then (3.16) becomes
\begin{eqnarray*}\hspace{1cm}& &\sum_{i\in\mbb{Z}}[(i+2)(i+1)-2]b_{i+2}(y+\be)^i
 =\frac{\al^2}{(y+\be)^4}+
\frac{2\al'}{(y+\be)^2}+\frac{4{\be'}'}{y+\be}
\\ &&+2\al\gm(y+\be)+10\be'\gm(y+\be)^2 +2\gm'
(y+\be)^3+\gm^2(y+\be)^6.\hspace{2.9cm}(3.18)\end{eqnarray*} Thus
$$b_{-2}=\frac{\al^2}{4},\;\;b_0=-\al',\;\;b_1=-2{\be'}',\;\;b_3=\frac{\al\gm}{2},\eqno(3.19)$$
$$b_4=\be'\gm,\;\;b_5=\frac{\gm'}{9},\;\;b_8=\frac{\gm^2}{54},\eqno(3.20)$$
$$(i+3)ib_{i+2}=0\qquad\for\;\;i\neq
-4,-2,-1,1,2,3,6.\eqno(3.21)$$ Therefore,
\begin{eqnarray*}\hspace{1cm}
f&=&\frac{\al^2}{4(y+\be)^2}+\frac{\sgm}{y+\be}-\al'-2{\be'}'(y+\be)+\rho(y+be)^2
\\ & &+\frac{\al\gm}{2}(y+\be)^3+\be'\gm(y+\be)^4 +\frac{\gm'}{9}(y+\be)^5+
\frac{\gm^2}{54}(y+\be)^8,\hspace{2.6cm}(3.22)\end{eqnarray*}
where $\sgm$ and $\rho$ are arbitrary functions of $t$. \psp

{\bf Theorem 3.1}. {\it We have the following solution of the
equation (1.1) blowing up on the surface $y+\be(t)=0$:
\begin{eqnarray*}\hspace{1cm}
u&=&\frac{x^2}{(y+\be)^2}+\frac{\al x}{(y+\be)^2}
+\frac{2\be'x}{y+\be}+\gm (y+\be)^3x+\frac{\al^2}{4(y+\be)^2}\\ &
&+\frac{\sgm}{y+\be}-\al' -2{\be'}'(y+\be)+\rho(y+be)^2
+\frac{\al\gm}{2}(y+\be)^3\\ &&+\be'\gm(y+\be)^4
+\frac{\gm'}{9}(y+\be)^5+
\frac{\gm^2}{54}(y+\be)^8,\hspace{5.2cm}(3.23)\end{eqnarray*}
where $\al,\be,\gm,\sgm$ and $\rho$ are arbitrary functions of
$t$, whose derivatives appeared in the above exist in a certain
open set of $\mbb{R}$.}\psp

When $\al=\gm=\sgm=\rho=0$, the above solution becomes
$$u=\frac{x^2}{(y+\be)^2}+\frac{2\be'x}{y+\be}-2{\be'}'(y+\be).\eqno(3.24)$$
\pse

 Take the trivial solution $\xi=0$ of (3.5), which is the only
solution polynomial in $y$. Then (3.6) and (3.7) become
$$g_{yy}=0,\qquad f_{yy}=2g_t+g^2.\eqno(3.25)$$
Thus
$$g=\al(t)+\be(t)y.\eqno(3.26)$$
Hence
$$f_{yy}=\al^2+2\al'+2(\be'+\al\be)y+\be^2 y^2.\eqno(3.27)$$
So
$$f=\gm+\sgm
y+\frac{\al^2+2\al'}{2}y^2+\frac{\be'+\al\be}{3}y^3+\frac{\be^2}{12}
y^4,\eqno(3.28)$$ where $\gm$ and $\sgm$ are arbitrary functions
of $t$.\psp

{\bf Theorem 3.2}. {\it The following is a solution of the
equation (1.2):
$$u=(\al+\be y)x+\gm+\sgm
y+\frac{\al^2+2\al'}{2}y^2+\frac{\be'+\al\be}{3}y^3+\frac{\be^2}{12}
y^4,\eqno(3.29)$$ where $\al,\be,\gm$ and $\sgm$ are arbitrary
functions of $t$, whose derivatives appeared in the above exist in
a certain open set of $\mbb{R}$. Moreover, any solution polynomial
in $x$ and $y$ of (1.2) must be of the above form. The above
solution is smooth (analytic) if all $\al,\be,\gm$ and $\sgm$ are
smooth (analytic) functions of $t$. }\psp

{\bf Remark 3.3}. In addition to the solutions in Theorems 3.1 and
3.2, the equation (1.2) has the following simple solution:
$$u=\wp_\iota (y)\:x^2,\eqno(3.30)$$
where $\wp_\iota(w)$ is the Weierstrass's elliptic function
satisfying (2.40).

\section{3-D Khokhlov-Zabolotskaya Equation}

By comparing the terms of highest degree, we find that
 a  solution polynomial in $x$ of the equation (1.3) must be of the form:
$$u=f(t,y,z)+g(t,y,z)x+\xi(t,y,z)x^2,\eqno(4.1)$$
 where $f(t,y,z),\;g(t,y,z)$ and $\xi(t,y,z)$ are
suitably-differentiable functions to be determined. As
(3.2)-(3.7), the equation (1.3) is equivalent to:
$$\xi_{yy}+\xi_{zz}=6\xi^2,\eqno(4.2)$$
$$g_{yy}+g_{zz}-6g\xi=4\xi_t,\eqno(4.3)$$
$$f_{yy}+f_{zz}-2f\xi=2g_t+g^2.\eqno(4.4)$$

First we observe that
$$\xi=\frac{1}{(y\cos \al(t)+z\sin
\al(t)+\be(t))^2}\eqno(4.5)$$ is a solution of the equation (4.2),
where $\al$ and $\be$ are suitable differentiable functions of
$t$. With the above $\xi$, (4.3) becomes
$$g_{yy}+g_{zz}-\frac{6g}{(y\cos \al(t)+z\sin
\al(t)+\be(t))^2}= -\frac{8(\al'(-y\sin
\al+z\cos\al)+\be')}{(y\cos \al+z\sin \al+\be)^3}.\eqno(4.6)$$ In
order to solve (4.6), we change variables:
$$\zeta=\cos \al\:y+\sin
\al\:z+\be,\;\;\eta=-\sin \al\:y+\cos\al\:z.\eqno(4.7)$$ Then
$$\ptl_y=\cos \al\:\ptl_\zeta-\sin
\al\:\ptl_\eta,\qquad\ptl_z=\sin
\al\:\ptl_\zeta+\cos\al\:\ptl_\eta.\eqno(4.8)$$ Thus
$$\ptl_y^2+\ptl_z^2=(\cos \al\:\ptl_\zeta-\sin
\al\:\ptl_\eta)^2+(\sin
\al\:\ptl_\zeta+\cos\al\:\ptl_\eta)^2=\ptl_\zeta^2+\ptl_\eta^2.\eqno(4.9)$$
Note
$$\ptl_t(\zeta)=\al'\eta+\be',\qquad\ptl_t(\eta)=\al'(\be-\zeta).\eqno(4.10)$$

The equation (4.6) can be rewritten as:
$$g_{\zeta\zeta}+g_{\eta\eta}-6\zeta^{-2}g
=-8(\al'\eta+\be')\zeta^{-3}.\eqno(4.11)$$ In order to solve the
above equation, we assume
$$g=\sum_{i\in\mbb{Z}}a_i(t,\eta)\zeta^i.\eqno(4.12)$$
Now (4.11) becomes
$$\sum_{i\in\mbb{Z}}[((i+2)(i+1)-6)a_{i+2}+a_{i\eta\eta}]=
-8(\al'\eta+\be')\zeta^{-3},\eqno(4.13)$$ which is equivalent to
$$-4a_{-1}+a_{-3\eta\eta}=-8(\al'\eta+\be'),\;\;
(i+4)(i-1)a_{i+2}+a_{i\eta\eta}=0\qquad\for \;\;-3\neq
i\in\mbb{Z}.\eqno(4.14)$$ Hence
$$a_{-1}=\frac{1}{4}a_{-3\eta\eta}+2(\al'\eta+\be'),\;\;(i+4)(i-1)a_{i+2}=-a_{i\eta\eta}
\qquad\for\;\;-3\neq i\in\mbb{Z}.\eqno(4.15)$$ When $i=-4$ and
$i=1$, we get $a_{-4\eta\eta}=a_{1\eta\eta}=0$. Moreover, $a_{-2}$
and $a_3$ can be any functions.

 Take
$$a_3=\sgm,\;\;
a_{-2}=\rho,\;\;a_{-1}=2(\al'\eta+\be'),\eqno(4.16)$$
$$a_1=a_{-1-2i}=a_{-2-2i}=0\qquad\for\;\;0<i\in\mbb{Z}\eqno(4.17)$$
in order to avoid infinite number of negative powers of $\zeta$ in
(4.12), where $\sgm$ and $\rho$ are arbitrary functions of $t$ and
$\eta$ differentiable in a certain domain. By (4.15),
$$a_{3+2k}=\frac{(-1)^k\ptl_{\eta}^{2k}(\sgm)}{\prod_{i=1}^k(2i+5)(2i)}
=\frac{(-1)^k15\ptl_{\eta}^{2k}(\sgm)}
{(2k+5)(2k+3)(2k+1)!},\eqno(4.18)$$
$$a_{-2+2k}=\frac{(-1)^k\ptl_{\eta}^{2k}(\rho)}{\prod_{i=1}^k(2i)(2i-5)}=
\frac{(-1)^k(2k-1)(2k-3)\ptl_{\eta}^{2k}(\rho)}
{3(2k)!}.\eqno(4.19)$$ Therefore,
\begin{eqnarray*}\hspace{1.5cm}g&=&2(\al'\eta+\be')\zeta^{-1}+\sum_{k=0}^\infty(-1)^k
[\frac{15\ptl_{\eta}^{2k}(\sgm)\zeta^3} {(2k+5)(2k+3)(2k+1)!}\\ &
& +\frac{(2k-1)(2k-3)\ptl_{\eta}^{2k}(\rho)\zeta^{-2}}
{3(2k)!}]\zeta^{2k}\hspace{6.4cm}(4.20)\end{eqnarray*} is a
solution of (4.11).

By (4.9), (4.4) is equivalent to
$$f_{\zeta\zeta}+f_{\eta\eta}-2\zeta^{-2}f=2g_t+g^2.\eqno(4.21)$$
 Note
\begin{eqnarray*}& &g_t=2({\al'}'\eta+{\be'}'+(\al')^2\be)\zeta^{-1}-2(\al')^2-2(\al'\eta+\be')^2
\zeta^{-2} +\sum_{k=0}^\infty(-1)^k\zeta^{2k}\{\\ & &\times
\left(\frac{15\ptl_{\eta}^{2k}(\sgm_t+\al'(\be-\zeta)\sgm_\eta)\zeta^3}
{(2k+5)(2k+3)(2k+1)!}+\frac{(2k-1)(2k-3)\ptl_{\eta}^{2k}(\rho_t+\al'(\be-\zeta)\rho_\eta)\zeta^{-2}}
{3(2k)!}\right)\\
&&+(\al'\eta+\be') [\frac{15\ptl_{\eta}^{2k}(\sgm)\zeta^2}
{(2k+5)(2k+1)!}+\frac{(2k-1)(2k-2)(2k-3)\ptl_{\eta}^{2k}
(\rho)\zeta^{-3}}{3(2k)!}]\}.\hspace{1.7cm}(4.22)\end{eqnarray*}
For convenience of solving the equation (4.21), we denote
$$2g_t+g^2=\sum_{i=-4}^\infty
b_i(t,\eta)\zeta^i\eqno(4.23)$$ by (4.20) and (4.22). In
particular,
$$b_{-4}=\rho^2,\qquad b_{-3}=0,\eqno(4.24)$$
$$b_{-2}=2(\rho_t+\al'\be\rho_\eta)+\frac{\rho_{\eta\eta}\rho}{3},
\eqno(4.25)$$
$$b_{-1}=4[{\al'}'\eta+{\be'}'+(\al')^2\be]-2\al'\rho_\eta
+\frac{2}{3}(\al'\eta+\be')\rho_{\eta\eta},\eqno(4.26)$$
$$b_0=-4(\al')^2
+\frac{1}{3}(\rho_{t\eta\eta}+\al'\beta\rho_{\eta\eta\eta})+
\frac{1}{12}
\ptl^4_\eta(\rho)\rho+\frac{1}{36}\rho^2_{\eta\eta}.\eqno(4.27)$$

Suppose that
$$f=\sum_{i\in\mbb{Z}}c_i(t,\eta)\zeta^i\eqno(4.29)$$
is a solution (4.21). Then
$$\sum_{i\in\mbb{Z}}[((i+2)(i+1)-2)c_{i+2}+c_{i\eta\eta}]\zeta^i=\sum_{r=-4}^\infty
b_r\zeta^r,\eqno(4.30)$$ equivalently
$$(i+3)ic_{i+2}=b_i-c_{i\eta\eta},\;\;(r+3)rc_{r+2}=-c_{r\eta\eta},\qquad
r<-4\leq i.\eqno(4.31)$$ By the  above second equation , we take
$$c_r=0\qquad\for\;\;r<-4\eqno(4.32)$$
to avoid infinite number of negative powers of $\zeta$ in (4.29).
Letting $i=-3,0$, we get
$$b_{-3}=c_{-3\eta\eta},\qquad b_0=c_{0\eta\eta}.\eqno(4.33)$$

The first equation is naturally satisfied because
$c_{-3}=-c_{-5\eta\eta}/10=0$. Taking $i=-2,-4$ and $r=-6$ in
(4.31), we obtain
$$c_0=\frac{1}{2}c_{-2\eta\eta}-\frac{1}{2}b_{-2},\qquad c_{-2}=\frac{1}{4}b_{-4}.
\eqno(4.34)$$ So
$$c_0=\frac{1}{8}\ptl_{\eta}^2(b_{-4})-\frac{1}{2}b_{-2}.\eqno(4.35)$$
Thus we get a constraint:
$$b_0=\frac{1}{8}\ptl_{\eta}^4(b_{-4})-\frac{1}{2}\ptl_\eta^2(b_{-2}),\eqno(4.36)$$
equivalently,
\begin{eqnarray*} \hspace{2cm}& &-4(\al')^2
+\frac{1}{3}(\rho_{t\eta\eta}+\al'\beta\rho_{\eta\eta\eta})+
\frac{1}{12} \ptl^4_\eta(\rho)\rho+\frac{1}{36}\rho^2_{\eta\eta}\\
&=&\frac{1}{8}\ptl_{\eta}^4(\rho^2)-\rho_{t\eta\eta}-
\al'\be\rho_{\eta\eta\eta}-\frac{\ptl_\eta^2(\rho_{\eta\eta}\rho)}{6}
.\hspace{5.3cm}(4.37)\end{eqnarray*} Thus
$$96(\rho_{t\eta\eta}+\al'\beta\rho_{\eta\eta\eta})+6
\ptl^4_\eta(\rho)\rho+
2\rho^2_{\eta\eta}-9\ptl_{\eta}^4(\rho^2)+12\ptl_\eta^2
(\rho_{\eta\eta}\rho)=288(\al')^2.\eqno(4.38)$$

It can be proved by considering the terms of highest degree that
any solution  of (4.38) polynomial in $\eta$ must be of the form
 $$\rho=\gm_0(t)+\gm_1(t)\eta+\gm_2(t)\eta^2.\eqno(4.39)$$
Then (4.38) becomes
$$6\gm_2'-5\gm_2^2=9(\al')^2.\eqno(4.40)$$
So
$$\al'=\frac{\es}{3}\sqrt{6\gm_2'-5\gm_2^2}\Rightarrow\al=\frac{\es}{3}\int
\sqrt{6\gm_2'-5\gm_2^2}dt,\eqno(4.41)$$ where $\es=\pm 1$. Replace
$\be$ by $-\be$ if necessary, we can take $\es=1$.
 Under the assumption
(4.39),
$$g=\rho\zeta^{-2}+2(\al'\eta+\be')\zeta^{-1}+\frac{\gm_2}{6}+\sum_{k=0}^\infty(-1)^k
\frac{15\ptl_{\eta}^{2k}(\sgm)\zeta^{3+2k}}{(2k+5)(2k+3)(2k+1)!}\eqno(4.42)$$
and
$$b_{-2}=2(\rho_t+\al'\be\rho_\eta)+\frac{2}{3}\gm_2\rho,\eqno(4.43)$$
$$b_{-1}=4[{\al'}'\eta+{\be'}'+(\al')^2\be]-2\al'\rho_\eta
+\frac{4}{3}(\al'\eta+\be')\gm_2,\eqno(4.44)$$
$$b_0=-4(\al')^2
+\frac{2}{3}\gm_2'+\frac{\gm_2^2}{9}.\eqno(4.45)$$ Denote
$$\Psi_{\la\be,\rho,\sgm\ra}(t,\eta,\zeta)=\sum_{i=1}^\infty
b_i\zeta^i.\eqno(4.46)$$ For any real function $F(t,\eta)$
analytic at $\eta=\eta_0$, we define
$$F(t,\eta_0+\sqrt{-1}\zeta)=\sum_{r=0}^\infty\frac{\ptl^r_\eta(F)(t,\eta_0)}{r!}
(\sqrt{-1}\zeta)^r. \eqno(4.47)$$

Note
\begin{eqnarray*}& &\sum_{k=0}^\infty(-1)^k
\frac{15\ptl_{\eta}^{2k}(\sgm)\zeta^{3+2k}}{(2k+5)(2k+3)(2k+1)!}
=15\zeta^2\int_0^\zeta\left(\sum_{k=0}^\infty(-1)^k
\frac{\ptl_{\eta}^{2k}(\sgm)\tau_1^{2k}}{(2k+5)(2k+3)(2k)!}\right)d\tau_1\\
&=&15\int_0^\zeta\tau_2\int_0^{\tau_2}\left(\sum_{k=0}^\infty(-1)^k
\frac{\ptl_{\eta}^{2k}(\sgm)\tau_1^{2k}}{(2k+5)(2k)!}\right)d\tau_1\:d\tau_2\\
&=&15\zeta^{-2}\int_0^\zeta\tau_3\int_0^{\tau_3}\tau_2\int_0^{\tau_2}\left(\sum_{k=0}^\infty(-1)^k
\frac{\ptl_{\eta}^{2k}(\sgm)\tau_1^{2k}}{(2k)!}\right)d\tau_1\:d\tau_2\:d\tau_3\\
&=&\frac{15}{2}\zeta^{-2}\int_0^\zeta\tau_3\int_0^{\tau_3}\tau_2\int_0^{\tau_2}[\sgm(t,\eta+
\sqrt{-1}\tau_1)+\sgm(t,\eta-\sqrt{-1}\tau_1)]d\tau_1\:d\tau_2\:d\tau_3
,\hspace{1.3cm}(4.48)\end{eqnarray*}
\begin{eqnarray*}& &\ptl_t
\left[\sum_{k=0}^\infty(-1)^k
\frac{15\ptl_{\eta}^{2k}(\sgm)\zeta^{3+2k}}{(2k+5)(2k+3)(2k+1)!}\right]
\\ &=&
\sum_{k=0}^\infty(-1)^k
\frac{15\ptl_{\eta}^{2k}(\sgm_t)\zeta^{3+2k}}{(2k+5)(2k+3)(2k+1)!}+\al'\be\sum_{k=0}^\infty(-1)^k
\frac{15\ptl_{\eta}^{2k+1}(\sgm)\zeta^{3+2k}}{(2k+5)(2k+3)(2k+1)!}
\\ & &-\al'\sum_{k=0}^\infty(-1)^k
\frac{15\ptl_{\eta}^{2k+1}(\sgm)\zeta^{4+2k}}{(2k+5)(2k+3)(2k+1)!}+(\al'\eta+\be')\sum_{k=0}^\infty(-1)^k
\frac{15\ptl_{\eta}^{2k}(\sgm)\zeta^{2+2k}}{(2k+5)(2k+1)!}
\\ &=&
\frac{15}{2}\zeta^{-2}\int_0^\zeta\tau_3\int_0^{\tau_3}\tau_2\int_0^{\tau_2}[\sgm_t(t,\eta+
\sqrt{-1}\tau_1)+\sgm_t(t,\eta-\sqrt{-1}\tau_1)]d\tau_1\:d\tau_2\:d\tau_3
\\ &&+\frac{15\al'(\zeta-\be)}{2\zeta^2}\sqrt{-1}
\zeta^{-2}\int_0^\zeta\tau_2\int_0^{\tau_2}\tau_1[\sgm(t,\eta+
\sqrt{-1}\tau_1)-\sgm(t,\eta-\sqrt{-1}\tau_1)] d\tau_1\:d\tau_2\\
&&+\frac{15}{2}(\al'\eta+\be')
\zeta^{-3}\int_0^\zeta\tau_2^3\int_0^{\tau_2}[\sgm(t,\eta+
\sqrt{-1}\tau_1)+\sgm(t,\eta-\sqrt{-1}\tau_1)] d\tau_1\:d\tau_2.
\hspace{1cm}(4.49)\end{eqnarray*} Hence
\begin{eqnarray*}& &g=
\rho\zeta^{-2}+2(\al'\eta+\be')\zeta^{-1}+\frac{\gm_2}{6}
+\frac{15}{2}\zeta^{-2}\\ & &\times
\int_0^\zeta\tau_3\int_0^{\tau_3}\tau_2\int_0^{\tau_2}[\sgm(t,\eta+
\sqrt{-1}\tau_1)+\sgm(t,\eta-\sqrt{-1}\tau_1)]d\tau_1\:d\tau_2\:d\tau_3,
\hspace{2.2cm}(4.50)\end{eqnarray*} by (4.42) and (4.48).
According to (4.23) and (4.46), we have
\begin{eqnarray*}& &\Psi_{\la\be,\rho,\sgm\ra}(t,\eta,\zeta)=\\ & &
\frac{225}{4}\zeta^{-4}\left(\int_0^\zeta\tau_3\int_0^{\tau_3}\tau_2\int_0^{\tau_2}[\sgm(t,\eta+
\sqrt{-1}\tau_1)+\sgm(t,\eta-\sqrt{-1}\tau_1)]d\tau_1\:d\tau_2\:d\tau_3\right)^2
\hspace{7cm}\end{eqnarray*}
\begin{eqnarray*} &
&
+15\zeta^{-2}\int_0^\zeta\tau_3\int_0^{\tau_3}\tau_2\int_0^{\tau_2}[\sgm_t(t,\eta+
\sqrt{-1}\tau_1)+\sgm_t(t,\eta-\sqrt{-1}\tau_1)]d\tau_1\:d\tau_2\:d\tau_3
\\ &&+\frac{15\al'(\zeta-\be)}{\zeta^2}\sqrt{-1}
\int_0^\zeta\tau_2\int_0^{\tau_2}\tau_1[\sgm(t,\eta+
\sqrt{-1}\tau_1)-\sgm(t,\eta-\sqrt{-1}\tau_1)] d\tau_1\:d\tau_2\\
&&+15(\al'\eta+\be')
\zeta^{-3}\int_0^\zeta\tau_2^3\int_0^{\tau_2}[\sgm(t,\eta+
\sqrt{-1}\tau_1)+\sgm(t,\eta-\sqrt{-1}\tau_1)] d\tau_1\:d\tau_2\\
& &
+15\left(\int_0^\zeta\tau_3\int_0^{\tau_3}\tau_2\int_0^{\tau_2}[\sgm(t,\eta+
\sqrt{-1}\tau_1)+\sgm(t,\eta-\sqrt{-1}\tau_1)]d\tau_1\:d\tau_2\:d\tau_3\right)\\&&\times
\zeta^{-2}\left(\rho\zeta^{-2}+(\al'\eta+\be')\zeta^{-1}+\frac{\gm_2}{6}\right).
\hspace{7.8cm}(4.51)\end{eqnarray*}

Now
$$c_{-2}=\frac{\rho^2}{4}\eqno(4.52)$$
by (4.24) and (4.34). According to (4.31) with $i=-3,0$, $c_{-1}$
and $c_2$ can be arbitrary. For convenience, we redenote
$$ c_{-1}=\kappa(t,\eta),\qquad c_2=\omega(t,\eta).\eqno(4.53)$$
 Moreover, (4.24), (4.35) and (4.43) imply
$$c_0=\frac{\rho^2_\eta}{4}-\rho_t-\al'\be\rho_\eta+\frac{\gm_2\rho}{6}.\eqno(4.54)$$
Furthermore, (4.31) and (4.44) yield
$$c_1=\frac{\kappa_{\eta\eta}}{2}-
2({\al'}'\eta+{\be'}'+(\al')^2\be)+\al'\rho_\eta
-\frac{2}{3}(\al'\eta+\be')\gm_2.\eqno(4.55)$$  In addition,
(4.31) and (4.53) gave $$c_{2k+3}=
\frac{(-1)^{k+1}\ptl_\eta^{2k+4}(\kappa)}{2(k+2)(2k+2)!}+
\sum_{i=0}^k\frac{(-1)^{k-i}(i+1)(2i)!}{(k+2)(2k+2)!}\ptl_\eta^{2(k-i)}(b_{2i+1}),
\eqno(4.56)$$
$$c_{2k+4}=\frac{(-1)^{k+1}3\ptl_\eta^{2k+2}(\omega)}{(2k+5)(2k+3)!}
+\sum_{i=0}^k\frac{(-1)^{k-i}(2i+3)(2i+1)!}{(2k+5)(2k+3)!}\ptl_\eta^{2(k-i)}(b_{2i+2})
\eqno(4.57)$$ for $0\leq k\in\mbb{Z}$.

Set
\begin{eqnarray*}& &\Phi_{\la\be,\rho,\sgm,\kappa,\omega\ra}(t,\eta,\zeta)=
\kappa\zeta^{-1}+
\frac{\kappa_{\eta\eta}\zeta}{2}+\omega\zeta^2+\sum_{i=3}^\infty
c_i\zeta^i\\
&=&-\zeta\ptl_\zeta\zeta^{-1}
\left[\sum_{k=0}^\infty(-1)^k\frac{\ptl_\eta^{2k}(\kappa)\zeta^{2k}}{(2k)!}\right]
+\zeta^2\sum_{k=0}^\infty(-1)^k\frac{3\ptl_\eta^{2k}(\omega)\zeta^{2k}}{(2k+3)(2k+1)!}
\\& &+\sum_{k=0}^\infty
\sum_{i=0}^k\frac{(-1)^{k-i}(i+1)(2i)!}{(k+2)(2k+2)!}\ptl_\eta^{2(k-i)}(b_{2i+1})\zeta^{2k+3}
\\ & &+\sum_{k=0}^\infty\sum_{i=0}^k\frac{(-1)^{k-i}(2i+3)(2i+1)!}{(2k+5)(2k+3)!}
\ptl_\eta^{2(k-i)}(b_{2i+2})\zeta^{2k+4}.\hspace{4.3cm}(4.58)\end{eqnarray*}
Note
\begin{eqnarray*}\hspace{1cm} & &\zeta^2\sum_{k=0}^\infty(-1)^k\frac{3\ptl_\eta^{2k}(\omega)\zeta^{2k}}{(2k+3)(2k+1)!}
\\ &=&\frac{3}{2}\zeta^{-1}\int_0^\zeta\tau_2\int_0^{\tau_2}[\omega(t,\eta+\sqrt{-1}\tau_1)
+\omega(t,\eta-\sqrt{-1}\tau_1)]d\tau_1\:d\tau_2.\hspace{2.2cm}(4.59)\end{eqnarray*}
Moreover,
$$\Psi_{\la\be,\rho,\sgm\ra}(t,\eta,0)=0,\qquad b_i=
\frac{\ptl_\zeta^i(\Psi_{\la\be,\rho,\sgm\ra})(t,\eta,0)}{i!}\qquad\for\;\;0<i\in\mbb{Z}.
\eqno(4.60)$$ Thus
\begin{eqnarray*}& &\Phi_{\la\be,\rho,\sgm,\kappa,\omega\ra}(t,\eta,\zeta)=
\frac{3}{2}\zeta^{-1}\int_0^\zeta\tau_2\int_0^{\tau_2}[\omega(t,\eta+\sqrt{-1}\tau_1)
+\omega(t,\eta-\sqrt{-1}\tau_1)]d\tau_1\:d\tau_2\\ & &
-\frac{1}{2}\zeta\ptl_\zeta\zeta^{-1}[\kappa(t,\eta+\sqrt{-1}\tau_1)
+\kappa(t,\eta-\sqrt{-1}\tau_1)]
\\& &+\sum_{k=0}^\infty
\sum_{i=0}^k\frac{(-1)^{k-i}(i+1)\ptl_\eta^{2(k-i)}\ptl_\zeta^{2i+1}(\Psi_{\la\be,\rho,\sgm\ra})(t,\eta,0)}
{(2i+1)(k+2)(2k+2)!}\zeta^{2k+3}
\\ & &+\sum_{k=0}^\infty\sum_{i=0}^k\frac{(-1)^{k-i}(2i+3)
\ptl_\eta^{2(k-i)}\ptl_\zeta^{2i+2}(\Psi_{\la\be,\rho,\sgm\ra})(t,\eta,0)
}{(2i+2)(2k+5)(2k+3)!}
\zeta^{2k+4},\hspace{3.5cm}(3.61)\end{eqnarray*} in which the
summations are finite if $\sgm(t,\eta)$ is polynomial in $\eta$.
According to (4.52)-(4.58) and (4.61),
\begin{eqnarray*}f&=&\Phi_{\la\be,\rho,\sgm,\kappa,\omega\ra}(t,\eta,\zeta)+
\frac{\rho^2}{4}\zeta^{-2}+\frac{\rho^2_\eta}{4}-\rho_t-\al'\be\rho_\eta
+\frac{\gm_2\rho}{6}\\&&-[
2({\al'}'\eta+{\be'}'+(\al')^2\be)-\al'\rho_\eta
+\frac{2}{3}(\al'\eta+\be')\gm_2]\zeta.\hspace{4.8cm}(4.62)\end{eqnarray*}
\pse

{\bf Theorem 4.1}. {\it In terms of  the notions in (4.7), we have
the following solution of the equation (1.3) blowing up on the
hypersurface $\cos \al(t)\:y+\sin \al(t)\:z+\be(t)=0$ ($\zeta
=0$):
\begin{eqnarray*}u&=&x^2\zeta^{-2}+[\rho\zeta^{-2}+2(\al'\eta+\be')\zeta^{-1}+\frac{\gm_2}{6}
+\frac{15}{2}\zeta^{-2}\\ & &\times
\int_0^\zeta\tau_3\int_0^{\tau_3}\tau_2\int_0^{\tau_2}[\sgm(t,\eta+
\sqrt{-1}\tau_1)+\sgm(t,\eta-\sqrt{-1}\tau_1)]d\tau_1\:d\tau_2\:d\tau_3]x\\
&&+\Phi_{\la\be,\rho,\sgm,\kappa,\omega\ra}(t,\eta,\zeta)+
\frac{\rho^2}{4}\zeta^{-2}+\frac{\rho^2_\eta}{4}-\rho_t-\al'\be\rho_\eta
+\frac{\gm_2\rho}{6}\\&&-[
2({\al'}'\eta+{\be'}'+(\al')^2\be)-\al'\rho_\eta
+\frac{2}{3}(\al'\eta+\be')\gm_2]\zeta,\hspace{4.8cm}(4.63)\end{eqnarray*}
where the  involved parametric functions $\rho$ is given in
(4.39), $\al$ is given in (4.41) and $\be$ is any function of $t$.
Moreover, $\sgm,\;\kappa,\;\omega$ are real functions in real
variable $t$ and $\eta$, and
$\Phi_{\la\be,\rho,\sgm,\kappa,\omega\ra}(t,\eta,\zeta)$is given
in (4.61) via (4.51).}\psp

When $\sgm=\kappa=\omega=0$, the above solution becomes:
\begin{eqnarray*}u&=&x^2\zeta^{-2}+[\rho\zeta^{-2}+2(\al'\eta+\be')\zeta^{-1}
+\frac{\gm_2}{6}]x +
\frac{\rho^2}{4}\zeta^{-2}+\frac{\rho^2_\eta}{4}-\rho_t\\
& &-\al'\be\rho_\eta +\frac{\gm_2\rho}{6}-[
2({\al'}'\eta+{\be'}'+(\al')^2\be)-\al'\rho_\eta
+\frac{2}{3}(\al'\eta+\be')\gm_2]\zeta,\hspace{2cm}(4.64)\end{eqnarray*}

Next we consider $\xi=0$, which is the only solution polynomial in
$y$ and $z$ of (4.2). In this case, (4.3) and (4.4) becomes:
$$g_{yy}+g_{zz}=0,\qquad f_{yy}+f_{zz}=2g_t+g^2.\eqno(4.65)$$
The above first equation is classical two-dimensional Laplace
equation, whose solutions are called {\it harmonic functions}. In
order to find simpler expressions of  the solutions of the above
equations, we introduce a new notion. A complex function
$$G(\mu)\;\;\mbox{is
called {\it bar-homomorphic}
if}\;\;\ol{G(\mu)}=G(\ol{\mu}).\eqno(4.66)$$ For instance,
trigonometric functions, polynomials with real coefficients and
elliptic functions with bar-invariant periods are bar-homomorphic
functions. The extended function $F(t,\mu)$ in (4.47) is
bar-homomorphic in $\mu$.

As (4.20), it can be proved by power series that the general
solution of the first equation in (4.65) is:
$$g=(\sgm+\sqrt{-1}\rho)(t,y+\sqrt{-1}z)+(\sgm-\sqrt{-1}\rho)(t,y-\sqrt{-1}z),\eqno(4.67)$$
where $\sgm(t,\mu)$ and $\rho(t,\mu)$ are complex functions in
real variable $t$ and bar-homomorphic in complex variable $\mu$.
Set
$$w=y+\sqrt{-1}z,\qquad \ol{w}=y-\sqrt{-1}z.\eqno(4.68)$$
Then the Laplace operator
$$\ptl_y^2+\ptl_z^2=4\ptl_w\ptl_{\ol{w}}.\eqno(4.69)$$
The second equation in (4.65) is equivalent to:
\begin{eqnarray*}\hspace{1cm}&&\ptl_w\ptl_{\ol{w}}(f)=
\frac{g_t}{2}+\frac{g^2}{4}=\frac{1}{2}(\sgm_t+\sqrt{-1}\rho_t)(t,w)
+(\sgm_t-\sqrt{-1}\rho_t)(t,\ol{w})\\ &
&+\frac{1}{4}[(\sgm+\sqrt{-1}\rho)(t,w)+(\sgm-\sqrt{-1}\rho)(t,\ol{w})]^2.
\hspace{5.3cm}(4.70)\end{eqnarray*} Hence the general solution of
the second equation in (4.65) is:
\begin{eqnarray*}&&f=\int_{\ol{w_1}}^{\ol{w}}\int_{w_1}^{w}\{\frac{1}{2}
[(\sgm_t+\sqrt{-1}\rho_t)(t,\mu_1)
+(\sgm_t-\sqrt{-1}\rho_t)(t,\ol{\mu_1})]\\ &
&+\frac{1}{4}[(\sgm+\sqrt{-1}\rho)(t,\mu_1)+(\sgm-\sqrt{-1}\rho)(t,\ol{\mu_1})]^2\}d\mu_1\:d\ol{\mu_1}\\
& &+(\kappa+\sqrt{-1}\omega)(t,w)
+(\kappa-\sqrt{-1}\omega)(t,\ol{w}),\hspace{6.8cm}(4.71)\end{eqnarray*}
where $\kappa(t,\mu)$ and $\omega(t,\mu)$ are complex functions in
real variable $t$ and bar-homomorphic in complex variable $\mu$,
and $w_1$ is a complex constant.
 \psp

{\bf Theorem 4.3}. {\it In terms of the notions in (4.67), the
following is a solution polynomial in $x$ of the equation (1.3):
\begin{eqnarray*}&
&u=[(\sgm+\sqrt{-1}\rho)(t,w)+(\sgm-\sqrt{-1}\rho)(t,\ol{w})]x+
\int_{\ol{w_1}}^{\ol{w}}\int_{w_1}^{w}\{\frac{1}{2}[(\sgm_t+\sqrt{-1}\rho_t)(t,\mu_1)
\\ & &+(\sgm_t-\sqrt{-1}\rho_t)(t,\ol{\mu_1})]+\frac{1}{4}[(\sgm+\sqrt{-1}\rho)
(t,\mu_1)+(\sgm-\sqrt{-1}\rho)(t,\ol{\mu_1})]^2\}d\mu_1\:d\ol{\mu_1}\\
& & +(\kappa+\sqrt{-1}\omega)(t,w)
+(\kappa-\sqrt{-1}\omega)(t,\ol{w}),\hspace{6.8cm}(4.72)\end{eqnarray*}
 where
$\sgm(t,\mu),\;\rho(t,\mu),\;\kappa(t,\mu)$ and $\omega(t,\mu)$
are complex functions in real variable $t$ and bar-homomorphic in
complex variable $\mu$ (cf. (4.66). Moreover, the above solution
is smooth (analytic) if all $\sgm,\;\rho,\;\kappa$ and $\omega$
are smooth (analytic) functions. In particular, any solution of
the equation (1.3) polynomial in $x,y,z$ must be of the form
(4.72) in which $\sgm,\;\rho,\;\kappa$ and $\omega$ are polynomial
in $\mu$.}\psp

{\bf Remark 4.4}. In addition to the solutions in Theorems 4.1 and
4.2, the equation (1.3) has the following simple solution:
$$u=\wp_\iota(ay+bz)\:x^2,\eqno(4.73)$$
where $\wp_\iota(w)$ is the Weierstrass's elliptic function and
$a,b$ are real constants such that $a^2+b^2=1$.

\bibliographystyle{amsplain}

\begin{thebibliography}{10}

\bibitem{} M. Akiyama and T. Kamakura, Elliptically curved acoustic
lens emitting strongly forcused finite-amplitude beams: Application
of the spherical beam equation model to the theoretical prediction,
{\it Acous. Sci. Tech.} {\bf 26} (2005), 179-284.

\bibitem{} A. G. Bagdoev and L. G. Petrosyan, Justification of
the applicability of short wave equations in obtaining an equation
for modulation of a gas-fluid mixture, {\it Izv. Akad. Nauk
Armyan. SSR Ser. Mekh}. {\bf 38} (1985), no. 4, 58-66.

\bibitem{} S. Ermakov, Short wave/long wave interaction and
amplification of decimeter-scale wind waves in film slicks, {\it
Geophysical Research  Absstracts} {\bf 8} (2006), 00469.


\bibitem{} J. Gibbons, The Khokhlov-Zabolotskaya equation and the inverse
scattering problem of classical mechanics, {\it Dynamical Problems
in Soliton Systems (Kyoto, 1984)}, 36-41, Springer, Berlin, 1985.

\bibitem{} N. H. Ibragimov, {\it Lie Group Analysis of Differential
Equations}, Volume 1, CRC Handbook, CRC Press, 1995.


\bibitem{} R. S. Khamitova, Group structure and a basis of conservation laws,
{\it Teor. Mat. Fiz} {\bf 52} (1982), no. 2, 244


\bibitem{} R. V. Khokhlov and E. A. Zabolotskaya, Quasi-plane waves in
nonlinear acoustics of bounded bundles, {\it Akust. Zh.} {\bf 15}
(1969), no. 1, 40


\bibitem{} S. A. Khristianovich and O. S. Razhov, On nonlinear reflection of
weak shock waves, {\it Prikl. Mat. Tekh.} {\bf 22} (1958), no. 5,
586

\bibitem{} A. Kocdryavtsev and V. Sapozhnikov, Symmetries of the
generalized Khokhlov-Zabolotskaya equation, {\it Acous. Phys.} {\bf
4} (1998), 541-546.

\bibitem{} S. Koshvaga, N. Makavarests, V. Grimalsky, A. Kotsarenko
and R. Enriquez, Spectrum of the sismic-electromagneic and acoustic
wave caused by seismic and volcano activity, {\it Natural Hazards
and Earth System Sciences} {\bf 5} (2005), 203-209.

\bibitem{} I. Kostin and G. Panasenko,  Khokhlov-Zabolotskaya-Kuzentsov-type
equation: Nonlinear acoustics in heterogeneous media, {\it SIMA J.
Math.} {\bf 40} (2008), 699-715.

\bibitem{} S. V. Korsunskii, Self-similar solutions of two-dimensional
equations of Khokhlov-Zabolotskaya type, {\it Mat. Fiz. Nelinein.
Mekh.} {\bf 16} (1991), 81-87.


\bibitem{} R. Kraenkel, M. Manna and V. Merle,
Nonlinear short-wave propagation in ferrites, {\it Phys. Rev. E}
{\bf 61} (2000), 976-979.


\bibitem{} P. Kucharczyk, Group properties of the ``short waves" equations in
gas dynamics, {\it Bull. Acad. Polon. Sci., Ser. Sci. Techn.} {\bf
XIII} (1965), no. 5, 469


\bibitem{} B. A. Kupershmidt, Geometric-Hamiltonian forms for the
Kadomtsev-Petviashvili and Khokhlov-Zabolotskaya equations, {\it
Geometry in Partial Differential Equations}, 155-172, World
Scientific Publishing, River Edge, NJ, 1994.




\bibitem{} J. Lin and J. Zhang, Similarity reductions for the
Khokhlov-Zabolotskaya equation, {\it Comm. Theoret. Phys.} {\bf 24}
(1995), no. 1, 69-74.

\bibitem{} V. V. Lychagin, I. S. Krasil'shchik and A. M. Vinogradov, {\it
Introduction to Geometry of Nonlinear Differential Equations,}
Nauka, Moscow, 1986.


\bibitem{} F. Martinez-Moras and E. Ramos, Higher dimensional
classical W-algebras, {\it Commun. Math. Phys.} {\bf 157} (1993),
573-589.

\bibitem{} O. Morozov, Cartan's structure theory of symmetry
pseudo-groups for the Khokhlov-Zabolotskaya equation, {\it Acta
Appl. Math.} {\bf 101} (2008), 231-241.


\bibitem{} C. Roy and M. Nasker, Towards the conservation laws and Lie
symmetries for the Khokhlov-Zabolotskaya equation in three
dimensions, {\it J. Phys. A} {\bf 19} (1986), no. 10, 1775-1781.

\bibitem{} A. Rozanova, The Khokhlov-Zabolotskaya-Kuznetsov
equation, {\it Math. Acad. Sci. Paris} {\bf 344} (2007), 337-342.

\bibitem{} A. Rozanova Qualitative analysis of the Khokhlov-Zabolotskaya
equation, {\it Math. Models Mathods Appl. Sci.} {\bf 18} (2008),
781-812.


\bibitem{} S. Roy, C. Roy and M. De, Loop algebra of Lie symmetries for a
short-wave equation, {\it Internat. J. Theoret. Phys.} {\bf 27}
(1988), no. 1, 47-55.

\bibitem{} D. Sanchez, Long waves in ferromagnetic media, Khokhlov-Zabolotskaya
equation, {\it J. Diff. Equ.} {\bf 210} (2005), 263-289.


\bibitem{} F. Schwarz, Symmetries of the Khokhlov-Zabolotskaya equation, {\it
J. Phys. A} {\bf 20} (1987), no. 6, 1613.

\bibitem{} A. M. Vinogradov and E. M. Vorob'ev, Application of symmetries for
finding of exact solutions of Khokhlov-Zabolotskaya equation, {\it
Akust. Zh.} {\bf 22} (1976), no. 1, 22


\end{thebibliography}

\end{document}